\documentclass[12pt]{article}
\usepackage{amsmath,amsfonts,amssymb,amsthm,latexsym}
\usepackage[mathscr]{eucal}
\usepackage{graphicx} 
\usepackage{textcomp}
\newtheorem{definition}{Definition}
\newtheorem{prop}{Proposition}
\newtheorem{theorem}{Theorem}

\begin{document}

\centerline{\bf 1-Density Operators and Algebraic Version of The Hohenberg-Kohn Theorem }

\centerline {A. I. Panin}

\centerline{ \sl Chemistry Department, St.-Petersburg State University,}
\centerline {\sl University prospect 26, St.-Petersburg 198504, Russia }
\centerline { \sl e-mail: andrej@AP2707.spb.edu }

\sloppy

\bigbreak
{\bf ABSTRACT: }{\small
Interrelation of the Coleman's representabilty theory for 1-density operators and abstract algebraic form of the Hohenberg-Kohn theorem is studied in detail. Convenient realization of the Hohenberg-Kohn set of classes of 1-electron operators and the Coleman's set of ensemble representable 1-density operators  is presented.  Dependence of the Hohenberg-Kohn class structure on the boundary properties of the ground state 1-density operator is established and is illustrated on concrete simple examples. Algorithm of restoration of many electron determinant ensembles from a given 1-density diagonal is described. Complete description of the combinatorial structure of Coleman's polyhedrons is obtained.  }

\bigbreak

\bigbreak
\hrule
\bigbreak
{\bf I. Introduction}

In the middle of the twentieth century there appeared two papers  that became contemporary classics and called into being a new branch of quantum chemistry. The first paper, dedicated, at first glance, to solution of a very abstract problem of representability of 1-density operators by ensembles of many electron states, was written by Coleman and published in 1963 \cite{Coleman-1}. One year later, in 1964, Hohenberg and Kohn published their theorem \cite{Kohn-1} about correspondence between representable densities and external potentials. From mathematical point of view both results belong to the functional analysis and have close interrelation. Coleman's theorem gives explicit analytic description of the set of all ensemble representable 1-density operators and Hohenberg-Kohn theorem suggests implicit way of parametrization of representable 1-density functions by certain classes of 1-electron operators (potentials). Further development of Coleman's approach was mainly concentrated on attempts to get the necessary and sufficient conditions of ensemble representability for density operators of higher order, especially for 2-electron density operators (see \cite{Coleman-2, Coleman-3} and references therein). It turned out to be a very complicated mathematical problem and to the best of our knowledge the constructive analytic description of the set of all representable 2-density operators is still not found. On the contrary, mathematically almost trivial Hohenberg-Kohn  statement   became a banner of the density functional community and is considered at present as the foundation of the density functional theory (DFT) (see, e.g., \cite{Parr-Yang}-\cite{ Yang}).

The present paper is an attempt  to give reasonably rigorous analysis of both the Coleman's theory and abstract form of the Hohenberg-Kohn theorem for the case when one-electron sector of the Fock space is of finite dimension.

Section II is dedicated to 1-density operators and their properties. Besides formulation of different versions of the Coleman's theorem, a useful realization of the set of all ensemble representable 1-density operators is given. Combinatorial structure of Coleman's polyhedrons is discussed in Appendix A. In Appendix B an example of constructive description of the set of 1-density operators representable by pure states may be found. 

In Section III  general iteration formula for restoration of many electron determinant ensembles from some fixed representable 1-electron diagonal is discussed. This formula first appeared in our electronic publication \cite{Panin}.
Concrete numerical examples of using this formula are given.

In Section IV a very general algebraic formulation of the Hohenberg-Kohn statement is analyzed. Abstract considerations are accompanying by concrete examples demonstrating dependence of 1-density diagonal matrix elements on parametrized classes of 1-electron operators.

\bigbreak
\bigbreak

{\bf II. General Properties of 1-Density Operators}
\bigbreak

In coordinate representation density operator of the first order (1-density operator)  is introduced as the integral operator  
\begin{equation}
\rho:\psi(x)\to \int \rho(x,x')\psi(x')dx'
\label{coord}
\end{equation} 
acting on the one-electron sector of the Fock space. Here $x$ is a vector of space-spin variables and $\rho(x,x')$ is a kernel of the operator (\ref{coord}).  This operator is said to be representable by a pure $p-$electron state $\Psi(x_1,\ldots,x_p)$ if its kernel can be obtained   by contraction (integration) of the product $\Psi(x,\ldots,x_p)\Psi^*(x',\ldots,x_p)$ over $p-1$ variables: 
\begin{equation}
\rho(x,x')=\underbrace{\int\ldots\int}_{p-1}\Psi(x,x_2,\ldots,x_p)\Psi^*(x',x_2,\ldots,x_p)dx_2\ldots dx_p
\label{kernel}
\end{equation}
Contraction of an ensemble of arbitrary finite family of $p-$electron states
\begin{eqnarray}
\rho(x,x')=\sum\limits_{i=1}^{\kappa}\lambda_i\underbrace{\int\ldots\int}_{p-1}\Psi_i(x,x_2,\ldots,x_p)\Psi_i^*(x',x_2,\ldots,x_p)dx_2\ldots dx_p\\
\left (\lambda_i\ge 0,\qquad \sum\limits_{i=1}^{\kappa}\lambda_i=1\right )\qquad\qquad\qquad\qquad\nonumber
\end{eqnarray}
gives  a kernel of the so-called ensemble representable density operator. The diagonal part $\rho(x)= \rho(x,x)$ of this kernel is called (ensemble) representable density function and  plays fundamental role in the density functional theory.  

Directly from definition three properties of representable 1-density operators are easily deduced: 

(1) $\rho^{\dagger}=\rho$ (Hermiteancy); 

(2) $\rho\ge 0$ (positive semi-definiteness); 

(3) $Tr\ \rho=1$. 

Density operators satisfying conditions (1)-(3) are not necessarily representable. The necessary and sufficient conditions  of the ensemble representability of 1-density operators was found by Coleman \cite{Coleman-1,Coleman-2} and may be formulated as follows.
\begin{theorem}[Coleman] 
1-density operator $\rho$ is representable by an ensemble of $p-$electron states if and only if
\begin{eqnarray}
(1)\  &\rho^{\dagger}=\rho;\qquad\qquad\qquad\nonumber\\
(2)\  &0\le\rho\le\frac{1}{p}I;\qquad\qquad\qquad\\
(3)\  &Tr\ \rho=1.\qquad\qquad\qquad\nonumber
\end{eqnarray}
\label{coleman1}
\end{theorem}
Here $I$ is the 1-electron identity operator.
    
Now let us turn to the algebraic version of the representability problem confining ourselves to the finite-dimensional case. We suppose that 1-electron sector of the Fock space ${\cal F}_{N,1}$ is spanned by $n=|N|$ orthonormal molecular spin-orbitals (MSOs)  $\{\psi_i\}_{i\in N}$ and that standard creation-annihilation operators are associated with this MSO basis set. The electronic space corresponding to $p-$electron system is just the $p$th Grassmann power  ${\cal F}_{N,p}=\bigwedge^p{\cal F}_{N,1}$ of the 1-electron space. Following Dirac, we identify the algebra of linear operators over $p-$electron space ${\cal F}_{N,p}$ with the tensor product ${\cal F}_{N,p}\otimes {\cal F}^{\dagger}_{N,p}$ of this space and the space of $\frac{1}{2}$-linear functionals on ${\cal F}_{N,p}$. The contraction  operators is defined as 
\begin{equation}
c=\sum\limits_{i=1}^na_i\otimes a_i^{\dagger}.
\label{c}
\end{equation}
and it is easy to show that 1-density operator representable by a pure $p-$electron state $|\Psi\rangle$ is
\begin{equation}
\rho=\frac{1}{p!}c^{p-1}|\Psi\rangle\langle\Psi|
\label{purec}
\end{equation}   
Due to linearity of the contraction operator Eq.(\ref{purec}) is easily generalized to treat ensembles of $p$-electron states. The set of all representable by ensembles of $p$-electron states  1-density operators will be denoted by the symbol ${\EuScript V}_{N,p,1}$ and will be referred to as Coleman's set. It is a compact (and, consequently, closed) convex subset of the affine hyperplane $Tr\ \rho =1$ situated in the operator space ${\cal F}_{N,1}\otimes {\cal F}^{\dagger}_{N,1}$. From Coleman's theorem it readily follows that the set ${\EuScript V}_{N,p,1}$ is invariant with respect to transformations $u\otimes u^{\dagger}$ where $u\in {\sf U}({\cal F}_{N,1})$ is 1-electron unitary operator (the so-called unitary invariance of the representability problem). In other words, ${\EuScript V}_{N,p,1}$ is a ${\sf U}({\cal F}_{N,1})$-space with respect of 1-electron unitary group action (see, e.g. \cite{Rowe-1,Rowe-2}).  

The structure of the set ${\EuScript P}_{N,p,1}$ of 1-density operators representable by pure $p$-electron states is more complicated.  It is the image of the projective space ${\mathbb P}({\cal F}_{N,p})$ over the $p$-electron sector of the Fock space with respect to the contraction $\frac{1}{p!}c^{p-1}$. Since ${\mathbb P}({\cal F}_{N,p})$ is compact and connected space, its image with respect to continuous mapping (contraction) is also compact and connected. In addition, projective spaces do not admit global parametrizations being the simplest examples of the so-called analytic manifolds (see, e.g., \cite{Wells,Godbillon}). It may therefore  be expected that  global parametrization does not exist also  for the set ${\EuScript P}_{N,p,1}$ as a whole.        

In non-relativistic quantum chemistry molecular spin-orbitals are presented as (tensor) products of spatial and spin functions and it is normally assumed that $p$-electron states under consideration have  a fixed projection $M_S$ of the total spin. 1-electron sector of the Fock space can be decomposed into  a direct sum of its $\alpha$- and $\beta$-subspaces and 1-density operator becomes a direct sum of $\alpha$- and $\beta$-components. There are two equivalent forms of the representability conditions in this case. If the total 1-density operator is written as $\rho=\rho_{\alpha}\oplus \rho_{\beta}$ then $\rho$ is representable if and only if 
\begin{eqnarray}
(1)\  &\rho^{\dagger}_{\sigma}=\rho_{\sigma};\qquad\qquad\qquad\nonumber\\
(2)\  &0\le\rho_{\sigma}\le\frac{1}{p}I_{\sigma};\qquad\qquad\qquad\\
(3)\  &Tr\ \rho_{\sigma}=\frac{p_{\sigma}}{p}.\qquad\qquad\qquad\nonumber
\label{repab1}
\end{eqnarray}
If $\rho$ is written as $\rho=\frac{p_{\alpha}}{p}\rho_{\alpha}\oplus \frac{p_{\beta}}{p}\rho_{\beta}$ then it is representable if and only if 
\begin{eqnarray}
(1)\  &\rho^{\dagger}_{\sigma}=\rho_{\sigma};\qquad\qquad\qquad\nonumber\\
(2)\  &0\le\rho_{\sigma}\le\frac{1}{p_{\sigma}}I_{\sigma};\qquad\qquad\qquad\\
(3)\  &Tr\ \rho_{\sigma}=1.\qquad\qquad\qquad\nonumber
\label{repab2}
\end{eqnarray}
Here  $I_{\sigma}$ is the identity operator in $\sigma$-subspace of 1-electron space,  $p_{\sigma}$ is the number of $\sigma$-electrons, and $\sigma=\alpha, \beta$. In the second case $\rho_{\sigma}\in {\EuScript V}_{M,p_{\sigma},1}$ where $M$ is the orbital index set.     

Each 1-density operator is Hermitian and can therefore be diagonalized and its eigenfunctions (the so-called natural MSOs) constitute a basis of 1-electron sector of the Fock space and without loss of generality this basis can be considered to be orthonormal. In Dirac's notations spectral resolution of 1-density operator $\rho$ is 
\begin{equation}
\rho=\sum\limits_{i\in N}\lambda_i|\psi_i\rangle\langle\psi_i|
\label{sres}
\end{equation}
where $\lambda_i$ are the so-called natural occupancies. 
For 1-density operators of the form of Eq.(\ref{sres}) Coleman's theorem can be formulated as 
\begin{theorem}[Coleman] 
1-density operator $\rho$ presented as its spectral resolution is representable by an ensemble of $p$-electron states if and only if
\begin{eqnarray}
(1)&\lambda_i\in {\mathbb R}\quad\mbox{for all}\quad  i \in N\;\qquad\qquad\qquad\nonumber\\
(2)&0\le \lambda_i\le \frac{1}{p}\quad\mbox{for all}\quad  i \in N;\qquad\qquad\qquad\\
(3)&\sum\limits_{i\in N}\lambda_i=1.\qquad\qquad\qquad\nonumber
\end{eqnarray}
\label{coleman2}
\end{theorem}
In geometric terms for a fixed n-frame $\psi=(\psi_1,\ldots,\psi_n)$ (orthonormal ordered MSO basis) the set ${\sf V}^{\psi}_{N,p,1}$ of representable 1-density operators of the form (\ref{sres}) is the intersection of the standard simplex and a cube (with the edge length $\frac{1}{p}$) being therefore a convex polyhedron (situated in the hyperplane $\sum\limits_{i\in N}\lambda_i=1$). Both parametric and analytic descriptions of this polyhedron are available.

Parametric description: Polyhedron ${\sf V}^{\psi}_{N,p,1}$ is a convex hull of $\binom{n}p$ vertices 
\begin{equation}
v_{p\downarrow 1}(R)=\frac{1}{p}\sum\limits_{i\in R}|\psi_i\rangle\langle\psi_i|
\label{vertices}
\end{equation}   
where $R$ is $p$-element subset of the MSO index set $N$.

Analytic description: It is given by the Coleman's conditions (1)-(3) of Theorem \ref{coleman2}. In geometric terms the polyhedron ${\sf V}^{\psi}_{N,p,1}$ has $2n$ hyperfaces with normals
\begin{eqnarray}
{\bf n}_i^{0}&=&p|\psi_i\rangle\langle\psi_i|\\
{\bf n}_i^{1}&=&-p|\psi_i\rangle\langle\psi_i|+\sum\limits_{j\in N}|\psi_j\rangle\langle\psi_j|
\end{eqnarray}     
It is clear that for any fixed $n$-frame $\psi$ the polyhedron ${\sf V}^{\psi}_{N,p,1}$ is homeomorphic to the typical (standard) polyhedron ${\sf V}_{N,p,1}$ constituted by vectors $(\lambda_1,\ldots,\lambda_n)\in {\mathbb R}^n$ satisfying the Coleman's conditions (1)-(3) of Theorem \ref{coleman2}. Combinatorial structure of this polyhedron is outlined  in Appendix A.

In terms of natural occupancies it is easy to describe the border of the set ${\EuScript V}_{N,p,1}$: It is constituted by 1-density operators that have at least one natural occupancy equal to zero or $1/p$. For the border of  ${\EuScript V}_{N,p,1}$ standard symbol $\partial {\EuScript V}_{N,p,1}$ will be used. Interior part of ${\EuScript V}_{N,p,1}$ will be denoted as ${\overset{\circ}{\EuScript V}}_{N,p,1}$.  Note that the minimal number of non-vanishing natural occupancies is equal to the number of electrons $p$ and, if it is the case, then all these occupancies are equal to $1/p$ (single determinant density operators that can be considered as vertices of ${\EuScript V}_{N,p,1}$).  

Now we can describe two different realizations of  the set of ensemble representable 1-density operators. In the first realization $n$-frame $\psi$  is supposed to be fixed and any representable density operator is written in the form
\begin{equation}
\rho=\sum\limits_{i,j\in N}\rho_{ij}|\psi_i\rangle\langle\psi_j|
\end{equation}
where matrix $(\rho_{ij})$ satisfies conditions (1)-(3) of theorem 1. The set ${\EuScript V}_{N,p,1}$ is the image of the mapping
\begin{equation}
r:(\lambda,u)\to u\lambda u^{\dagger}
\end{equation}
where $\lambda\in {\sf V}^{\psi}_{N,p,1}$ and $u$ is 1-electron unitary transformation. It is easy to see that this mapping is not injective. In this realization ${\EuScript V}_{N,p,1}$ is a union of orbits with respect to the unitary group action where for each orbit its representative, diagonal in the basis $\psi$,  is selected.   

In the second realization the set of ensemble representable 1-density operators is considered as a disjoint union (sum) of
fibres  ${\sf V}^{\psi}_{N,p,1}$
\begin{equation}
{\widetilde {\EuScript V}_{N,p,1}}=\bigsqcup\limits_{\psi \in {\EuScript N}}{\sf V}^{\psi}_{N,p,1}
\end{equation}   
where ${\EuScript N}$ is the manifold of all orthonormal $n$-frames of 1-electron space ${\cal F}_{N,1}$. In more formal terms ${\widetilde {\EuScript V}_{N,p,1}}$ is a total space of (trivial) fibre bundle with the manifold ${\EuScript N}$ as its base and the polyhedron ${\sf V}_{N,p,1}$ as its typical fibre (see, e.g., \cite{Wells, Godbillon}). In such a realization the set of ensemble representable 1-density operators is homeomorphic to the Cartesian product ${\EuScript N}\times {\sf V}_{N,p,1}$. 1-electron unitary group acts transitively on the base ${\EuScript N}$ as
\begin{equation}
\psi'=u\psi=(\psi_1,\ldots,\psi_n)U
\end{equation}
where $U$ is a unitary matrix connecting two $n$-frames
\begin{equation}
\psi'_i=u\psi_i=\sum\limits_{j=1}^n\psi_jU_{ji},
\end{equation} 
and the induced mapping $u\otimes u^{\dagger}$ maps fibre ${\sf V}^{\psi}_{N,p,1}$ onto fibre ${\sf V}^{\psi'}_{N,p,1}$.

The set of 1-density operators representable by pure $p$-electron states can also be considered as a disjoint union of fibres  
\begin{equation}
{\widetilde {\EuScript P}_{N,p,1}}=\bigsqcup\limits_{\psi \in {\EuScript N}}{\sf P}^{\psi}_{N,p,1}
\label{pure}
\end{equation}   
where for each $\psi \in {\EuScript N}$ the fibre ${\sf P}^{\psi}_{N,p,1}$ is a compact connected subset of ${\sf V}^{\psi}_{N,p,1}$ necessarily containing all its vertices and the central point $\frac{1}{n}\sum\limits_{i=1}^n|\psi_i\rangle\langle\psi_i|$. An example of explicit description of a fibre ${\sf P}^{\psi}_{N,p,1}$ is given in Appendix B. It is pertinent to note that implicitly fibre bundles appeared in quantum chemistry  at its infancy in disguise of the so-called multi-configuration self-consistent field (MCSCF) theory. MCSCF fibre bundles have the set  ${\EuScript N}$ of $n$-frames as their base (SCF part) and unit spheres of some fixed dimension as the typical fibres (MC part). 
  
If $n$-frame $\psi=(\psi_1,\ldots,\psi_n)$ is fixed then for any integer $q=1,2,\ldots, n$ it is possible to define the diagonal mapping
\begin{equation}
d_{\psi}:\sum\limits_{R,S\subset  N}^{(q)}{\Lambda}_{RS}|R\rangle\langle S|\to \sum\limits_{R\in N}^{(q)}{\Lambda}_{RR}|R\rangle\langle R|
\label{dmap}
\end{equation}
where symbol $|R\rangle$ stands for $q$-electron determinant built on spin-orbitals $\psi_1,\ldots,\psi_n$ with indices from subset $R$. If $R=\{r_1<\ldots < r_q\}$ then $|R\rangle$ may be interpreted as either a factorable $p$-vector $\psi_{r_1}\wedge\ldots\wedge\psi_{r_q}$ in Grassmann algebra (electronic Fock space)  or as a vector $a^{\dagger}_{r_1}\ldots a^{\dagger}_{r_q}|\emptyset \rangle$ obtained by successive application of the fermion creation operators associated with a given MSO basis set. 
\begin{definition}
Vector $\lambda\in \mathbb R^n$ is called representable if it can be realized as a diagonal of  1-density operator with respect to some fixed basis.
 \end{definition}
It is easy to see that  the diagonal mapping (\ref{dmap}) commute with the contraction operator. This means, in particular, that the diagonal of any representable 1-density operator is representable by an ensemble of determinant states and  $d_{\psi}\left ({\EuScript V}_{N,p,1} \right )={\sf V}^{\psi}_{N,p,1}$. Of course,  $d_{\psi}(\rho)\in {\sf V}^{\psi}_{N,p,1}$ does not imply $\rho \in  {\EuScript V}_{N,p,1}$.
However, the following statement holds true.
\begin{prop} $d_{\psi}(\rho)\in {\sf V}^{\psi}_{N,p,1}$ implies the existence of representable by a pure state density operator $\rho'$ such that $d_{\psi}(\rho)=d_{\psi}(\rho')$
\end{prop}
\textbf{Proof.}
If $d_{\psi}(\rho)\in {\sf V}^{\psi}_{N,p,1}$ then, by definition, there exists an ensemble of $p$-electron determinant states
\begin{equation}
\Lambda=\sum\limits_{R\subset N}^{(p)}\Lambda_R|R\rangle\langle R|
\end{equation}
such that $d_{\psi}(\rho)=\frac{1}{p!}c^{p-1}\Lambda$, and any $p$-electron wave function of the form 
\begin{equation}
\Psi(\theta)=\sum\limits_{R\subset N}^{(p)}\exp(i\theta_R)\Lambda_R^{\frac{1}{2}}|R\rangle
\label{branches}
\end{equation}
corresponds to the pure state $|\Psi(\theta)\rangle\langle\Psi(\theta)|$ such that its contraction gives representable 1-density operator $\rho '$ with $d_{\psi}(\rho'(\theta))=d_{\psi}(\rho)$ for arbitrary phase vector $\theta$$\Box$ 

{\rm\bf Corollary.}

\begin{equation}
d_{\psi}\left ({\EuScript P}_{N,p,1}\right )={\sf V}^{\psi}_{N,p,1}
\end{equation}

The connection between two aforementioned realizations of the set of the ensemble representable 1-density operators is established by the  following simple assertion.

\begin{prop}  

{\rm (1)} {\it $\rho\in {\EuScript V}_{N,p,1}$ if and only if $d_{\psi}(\rho)\in {\sf V}^{\psi}_{N,p,1}$ for any $n$-frame $\psi$;}  

{\rm (2)}  {$\rho\in \partial {\EuScript V}_{N,p,1}$ if and only if there exists $n$-frame $\psi$ such that $d_{\psi}(\rho)\in \partial {\sf V}^{\psi}_{N,p,1}$;}

{\rm (3)} {\it $\rho\in {\overset{\circ}{\EuScript V}}{\vphantom V}^{\psi}_{N,p,1}$ if and only if $d_{\psi}(\rho)\in {\overset{\circ}{\sf V}}_{N,p,1}$ for any $n$-frame $\psi$.}

\end{prop}

\textbf{Proof.} Statement (1) readily follows from the Coleman's theorem. As for statement (2), it is sufficient to prove that $d_{\psi}(\rho)\in \partial {\sf V}^{\psi}_{N,p,1}$ implies $\rho\in \partial {\EuScript V}_{N,p,1}$. Indeed,  if $\rho\in \partial {\EuScript V}_{N,p,1}$ then in the basis $\psi$ of natural MSOs $d_{\psi}(\rho)\in \partial {\sf V}^{\psi}_{N,p,1}$. Let us suppose therefore that $d_{\psi}(\rho)\in \partial {\sf V}^{\psi}_{N,p,1}$ which means that $\rho_{ii}=0$ or $\rho_{ii}=\frac{1}{p}$ in the MSO basis under consideration. If $\left (U_{ji}\right )$ is  the matrix of coefficients of transformation from this basis to the basis of 1-density operator $\rho$ natural MSOs, then we have  
$$\rho_{ii}=\sum\limits_{j=1}^n |U_{ji}|^2\lambda_j$$
where $\lambda_i$ are natural occupancies. If $\rho_{ii}=0$ then from this equality it follows that there exists at least one index $j$ such that $\lambda_j=0$. If 
$\rho_{ii}=\frac{1}{p}$ then this equality may be recast as  
$$0=\sum\limits_{j=1}^n |U_{ji}|^2\left(\frac{1}{p}-\lambda_j\right )$$
and, consequently, there exists index $j$ such that $\frac{1}{p}-\lambda_{j}=0$. 

If $d_{\psi}(\rho)\in \rho\in {\overset{\circ}{\EuScript V}}_{N,p,1}$ for any $n$-frame $\psi$ then among natural occupancies of $\rho$ there are no occupancies equal to 0 or $\frac{1}{p}$. Assumption that $\rho\in {\overset{\circ}{\sf V}}{\vphantom V}^{\psi}_{N,p,1}$ and there exist $n$-frame $\psi$ such that $d_{\psi}(\rho)\in \partial {\sf V}^{\psi}_{N,p,1}$ contradicts to already proved assertion (2) of this Proposition      
\ $\Box$        

Explicit description of the border of the polyhedron ${\sf V}_{N,p,1}$ is given in Appendix A. Note that $d_{\psi}(\rho)\in {\overset{\circ}{\sf V}}{\vphantom V}^{\psi}_{N,p,1}$ for some $n$-frame $\psi$ does not imply $\rho\in {\overset{\circ}{\EuScript V}}_{N,p,1}$.

 Iteration formula for  generation of all ensembles of determinant states that are contracted into a given diagonal is obtained in the next section.    

\bigbreak

\bigbreak
{\bf III. Restoration of Many Electron Determinant Ensembles from Diagonal of One-Electron Density Matrix}
\bigbreak

With arbitrary vector ${\lambda}^{(0)} \in  {\sf V}^{\psi}_{N,p,1}$ it is
convenient to associate three index sets:
\begin{eqnarray}
Ind({\lambda}^{(0)})&=&\{i\in N: {\lambda}^{(0)}_i>0\}\\
Ind_{\frac{1}{p}}({\lambda}^{(0)})&=&\{i\in N: {\lambda}^{(0)}_i=\frac{1}{p}\}\\
Ind_a(\lambda^{(0)})&=&Ind(\lambda^{(0)})\backslash Ind_{\frac{1}{p}}(\lambda^{(0)})
\end{eqnarray}
Indices belonging to the last set will be called active.

Let us present vector ${\lambda}^{(0)} \in  {\sf V}^{\psi}_{N,p,1}$ as the convex combination
\begin{equation}
{\lambda}^{(0)} =p{\mu}^{R_0}v_{p\downarrow 1}(R_0)+(1-p{\mu}^{R_0}){\lambda}^{(1)}
\label{lambda0_1}
\end{equation}
where vertex $v_{p\downarrow 1}(R_0)$ is defined by  Eq.(\ref{vertices}),
\begin{equation}
{\lambda}^{(1)}=\sum\limits_{i \in R_0}\frac{{\lambda}^{(0)}_i-{\mu}^{R_0}}
{1-p{\mu}^{R_0}}|\psi_i\rangle\langle\psi_i|+\sum\limits_{i \in N\backslash R_0}
\frac{{\lambda}^{(0)}_i}{1-p{\mu}^{R_0}}|\psi_i\rangle\langle\psi_i|,
\label{res}
\end{equation}
and require the residual
vector  ${\lambda}^{(1)}$ to be representable. This requirement imposes
the following restrictions on the admissible values of parameter ${\mu}^{R_0}$:
\begin{equation}
\begin{cases}
0\le \frac{{\lambda}^{(0)}_i-{\mu}^{R_0}}{1-p{\mu}^{R_0}}\le \frac{1}{p},\  i\in R_0\\
0\le \frac{{\lambda}^{(0)}_i}{1-p{\mu}^{R_0}}\le \frac{1}{p},\ i\in N\backslash R_0\\
\end{cases}
\label{sys1}
\end{equation}
The solution  of this system  is the interval $\left [0,b^{R_0}\right ]\subset \mathbb R$ where 
\begin{equation}
b^{R_0}=\min\{\min_{i\in R_0}\{{\lambda}^{(0)}_i\},
\min_{i\in N \backslash
R_0}\{\frac{1}{p}-{\lambda}^{(0)}_i\}\}.
\label{b}
\end{equation}
If ${\mu}^{R_0}\in \left (0,b^{R_0}\right ]$ then we arrive at a non-trivial representation of
${\lambda}^{(0)}$  as a convex combination of vertex
$v_{p\downarrow 1}(R_0)$ and a certain representable residual
vector ${\lambda}^{(1)}$. 
\begin{prop}
Arbitrary vector $\lambda^{(0)}\in {\sf V}^{\psi}_{N,p,1}$ admits presentation in the form of Eq.(\ref{lambda0_1}) if and only if  
\begin{equation}
Ind_{\frac{1}{p}}(\lambda^{(0)})\subset R_0 \subset Ind(\lambda^{(0)})
\label{incl}
\end{equation}
\end{prop}
\textbf{Proof.} It is sufficient to show that the condition (\ref{incl}) is equivalent to existence of  non-vanishing boundary parameter $b^{R_0}$. But this follows directly from Eq.(\ref{b})$\Box$ 

Thus, with each $\lambda^{(0)}\in {\sf V}^{\psi}_{N,p,1}$ we can associate the set ${\cal P}(\lambda^{(0)})$ of $p-$element subsets satisfying the condition  (\ref{incl}). It is easy to show that 
\begin{equation}
|{\cal P}(\lambda^{(0)})|=\binom{{n_a(\lambda ^{(0)})}}{p-n_{\frac{1}{p}}(\lambda^{(0)})}
\end{equation}
where $n_a(\lambda^{(0)})=|Ind_a(\lambda^{(0)})|$ and $n_{\frac{1}{p}}(\lambda^{(0)})=|Ind_{\frac{1}{p}}(\lambda^{(0)})|$.
\begin{prop}
If  parameter $\mu^{R_0}$ in Eq.(\ref{lambda0_1}) is taken equal to its boundary value $b^{R_0}$ then $R_0\not\in {\cal P}(\lambda^{(1)})$.
\end{prop}
\textbf{Proof.} If $\mu^{R_0}=b^{R_0}=\frac{1}{p}$ then $Ind_{\frac{1}{p}}(\lambda^{(1)})=\emptyset$ and, obviously, $R_0\not\in {\cal P}(\lambda^{(1)})$. Let us suppose therefore that $0<b^{R_0}<\frac{1}{p}$. From Eq.(\ref{b}) it follows that there exists index $i_*\in Ind(\lambda^{(0)})$ such that either $i_* \in R_0$ and  $\mu^{R_0}=b^{R_0}=\lambda^{(0)}_{i_*}$ or $i_*\in N\backslash R_0$ and 
$\mu^{R_0}=b^{R_0}=\frac{1}{p}-\lambda^{(0)}_{i_*}$. From Eq.(\ref{res}) it is easy to see that in the first case $\lambda^{(1)}_{i_*}=0$ and, consequently, $i_*\in R_0$ but $i_*\not\in Ind(\lambda^{(1)})$. In the second case $\lambda^{(1)}_{i_*}=\frac{1}{p}$ which means that $i_* \in Ind_{\frac{1}{p}}(\lambda^{(1)})$ but $i_*\not\in R_0$.    
In both cases $R_0\not\in {\cal P}(\lambda^{(1)})$ $\Box$ 
\begin{prop}
If  parameter $\mu^{R_0}$ in Eq.(\ref{lambda0_1}) is taken equal to its boundary value $b^{R_0}$ and this boundary value is different from $ \frac{1}{p}$ then 
\begin{equation}
Ind_{\frac{1}{p}}(\lambda^{(0)})\subset Ind_{\frac{1}{p}}(\lambda^{(1)})
\label{incl1}
\end{equation}
\end{prop}
\textbf{Proof.} Let us show that $\lambda^{(0)}_{i}=\frac{1}{p}$ implies $\lambda^{(1)}_{i}=\frac{1}{p}$. Indeed, if     
$\mu^{R_0}=b^{R_0}={\lambda}^{(0)}_{i_*}<\frac{1}{p}$ and ${\lambda}^{(0)}_{i}=\frac{1}{p}$ for some $i\in R_0$ then 
${\lambda}^{(1)}_{i}=\frac{\frac{1}{p}-{\lambda}^{(0)}_{i_*}}{1-p{\lambda}^{(0)}_{i_*}}
=\frac{1}{p}$. If, on the other hand, $\mu^{R_0}=b^{R_0}=\frac{1}{p}-{\lambda}^{(0)}_{i_*}$ and ${\lambda}^{(0)}_{i}=\frac{1}{p}$ for some $i\in R_0$ then $1-p{\mu}^{R_0}=p{\lambda}^{(0)}_{i_*}$ and
${\lambda}^{(1)}_{i}=\frac{\frac{1}{p}-{\mu}^{R_0}}{p{\lambda}^{(0)}_{i_*}}
=\frac{1}{p}$ $\Box$ 

Iterating of Eq.(\ref{lambda0_1}) leads to the following expression
\begin{equation}
{\lambda}^{(0)}=\sum\limits_{i=0}^{k-1}\left [\prod\limits_{j=0}^{i-1}
(1-p{\mu}^{R_j})\right ]p{\mu}^{R_i}v_{p\downarrow 1}(R_i)+
\left[\prod\limits_{i=0}^{k-1}(1-p{\mu}^{R_i})\right ]{\lambda}^{(k)}
\label{iter}
\end{equation}
where
\begin{equation}
{\mu}^{R_i}\in \left(0,b^{R_i}\right ],
\label{muint}
\end{equation}
\begin{equation}
Ind_{\frac{1}{p}}(\lambda^{(k)})\subset R_i \subset Ind(\lambda^{(k)})
\label{filter}
\end{equation}
for $i=0,1,\ldots,k-1$.
Note that the residual vector in Eq.(\ref{iter}) is necessarily representable.
\begin{theorem}
For any vector ${\lambda}^{(0)} \in  V^{\psi}_{N,p,1}$
the residual vector in iteration formula (\ref{iter}) vanishes after
at most $n_a({\lambda}^{(0)})-1$ steps if at each step the boundary value of parameter ${\mu}^{R_i}$ is selected. 
\end{theorem}
\textbf{Proof.} Direct consequence of Propositions 1-3.

{\rm\bf Corollary 1.}
{\it The set $V^{\psi}_{N,p,1}$ is the
convex hull of $\binom{n}p$ vertices $v_{p\downarrow 1}(R)$.}

{\rm\bf Corollary 2.}

{\it Ensemble 
\begin{equation}
\Lambda(R_0,R_1,\ldots,R_{k_f}) =
\sum\limits_{i=0}^{k_f}\left [\prod\limits_{j=0}^{i-1}
(1-p{\mu}^{R_j})\right ]p{\mu}^{R_i}|R_i\rangle \langle R_i|
\end{equation}
of $p$-electron determinant states generated recurrently on the base of the iteration formula (\ref{iter}) with boundary values of parameters $\mu^{R_i}$ includes pairwise distinct $p$-element subsets and  
\begin{equation}
\frac{1}{p!}c^{p-1}\Lambda(R_0,R_1,\ldots,R_{k_f})= \lambda^{(0)}.
\end{equation}}

\begin{definition}
For ${\lambda}^{(0)} \in  V^{\psi}_{N,p,1}$ any its expansion corresponding to the boundary values of parameters $\mu^{R_i}$ will be called boundary expansions. 
\end{definition}

Let us consider simple example of boundary expansion. For $N=\{1,2,3,4\}$ and $p=2$ let us take representable rational vector $$\lambda^{(0)}=\left (\frac{7}{20},\frac{1}{4},\frac{1}{4},\frac{3}{20}\right )$$
and select $R_0=\{2,4\}$. With such a selection the boundary value  $b^{24}=\frac{3}{20}$ and Eq.(\ref{iter})
gives
$$\lambda^{(0)}=\frac{3}{10}v_{p\downarrow 1}(24)+\frac{7}{10}\lambda^{(1)}$$
where
$$\lambda^{(1)}=\left (\frac{1}{2},\frac{1}{7},\frac{5}{14},0\right ).$$
The next admissible subset $R_1$ may be chosen, say, as $R_1=\{1,2\}$. In this case $b^{12}=\frac{1}{7}$ and 
$$\lambda^{(1)}=\frac{2}{7}v_{p\downarrow 1}(12)+\frac{5}{7}\lambda^{(2)}$$
where
$$\lambda^{(2)}=\left (\frac{1}{2},0,\frac{1}{2},0\right )=v_{p\downarrow 1}(13).$$
The final expansion (which is by no means unique) is 
$$\lambda^{(0)}=\frac{3}{10}v_{p\downarrow 1}(24)+\frac{1}{2}v_{p\downarrow 1}(13)+\frac{1}{5}v_{p\downarrow 1}(12).$$
Note that if ${\lambda}^{(0)} \in  V^{\psi}_{N,p,1}$ is rational   then all intermediate 
vectors ${\lambda}^{(k)}$ in Eq.(\ref{iter}) are also rational if  at each iteration the boundary value of parameter ${\mu^{R_i}}$ is selected.       
  
Iteration formula (\ref{iter}) is valid for arbitrary choice of ${\mu}^{R_i}\in \left(0,b^{R_i}\right ]$ but Proposition 2 is not. In other words, the conditions (\ref{filter}) do not necessarily filter out $p-$element subsets selected on the previous steps. This may be used to construct expansions different from the boundary ones. And even more, it is easy to show that on the base of iteration formula (\ref{iter}) any pre-defined $p$-electron ensemble of determinant states may be obtained. 
\begin{prop}
Let 
\begin{equation}
\Lambda=\sum\limits_{R\subset N}^{(p)}|C_R|^2|R\rangle\langle R|.
\label{detens}
\end{equation} 
be an ensemble of $p$-electron determinant states and  $\lambda=\frac{1}{p!}c^{p-1}\Lambda$ is the corresponding 1-density diagonal with components  
\begin{equation}
\lambda_i=\frac{1}{p}\sum\limits_{R\ni i}^{(p)}|C_R|^2.
\label{components}
\end{equation}  
Then, using iteration formula (\ref{iter}), it is possible to restore the initial ensemble $\Lambda$.
\end{prop}
\textbf{Proof.} Let us suppose that $p$-element subsets in Eq.(\ref{detens}) corresponding to non-zero coefficients $|C_R|^2$ are ordered in some fixed (say, lexical) order. From Eq.(\ref{components}) it readily follows that
\begin{equation}
\frac{1}{p}|C_R|^2\le \lambda_i\nonumber
\end{equation}
for any $i\in R$. On the other hand, the equality $\lambda_i+\frac{1}{p}\sum\limits_{i\not\in R}|C_R|^2=\frac{1}{p}$ (that is a direct consequence of the normalization condition for $\Lambda$) implies that 
\begin{equation}
\frac{1}{p}|C_R|^2\le \frac{1}{p}-\lambda_i\nonumber
\end{equation}
for any $i\not\in R$. Thus, we can guarantee that $\mu^{R_0}=\frac{1}{p}|C_{R_0}|^2\in (0,b^{R_0}]$ and expansion  
Eq.(\ref{lambda0_1}) takes the form
\begin{equation}
\lambda=\lambda^{(0)}=|C_{R_0}|^2 v_{p\downarrow 1}(R_0)+(1-|C_{R_0}|^2)\lambda^{(1)}\nonumber
\end{equation} 
where $\lambda^{(1)}$ corresponds to the ensemble 
\begin{equation}
\Lambda^{(1)}=\frac{1}{1-|C_{R_0}|^2}\sum\limits_{\genfrac{}{}{0pt}{}{R\subset N}{(R\ne R_0)}}^{(p)}|C_R|^2|R\rangle\langle R|\nonumber
\end{equation}
Expanding  $\lambda^{(1)}$ in the form of Eq.(\ref{lambda0_1}) with $\mu^{R_1}=\frac{|C_{R_1}|^2}{p(1-|C_{R_0}|^2)}$, we come to the equality
\begin{equation}
\lambda^{(0)}=|C_{R_0}|^2 v_{p\downarrow 1}(R_0)+|C_{R_1}|^2 v_{p\downarrow 1}(R_1)+(1-|C_{R_0}|^2-|C_{R_1}|^2)\lambda^{(2)}\nonumber,
\end{equation} 
etc. On the pre-final step $k$ we have  the  ensemble
\begin{equation}
\Lambda^{(k-1)}= |C^{'}_{R_{k-1}}|^2|R_{k-1}\rangle\langle R_{k-1}|+ |C^{'}_{R_k}|^2|R_k\rangle\langle R_k|\nonumber
\end{equation}
where
\begin{equation}
|C^{'}_{R_{k-1}}|^2=\frac{|C_{R_{k-1}}|^2}{{1-\sum\limits_{i=0}^{k-2}|C_{R_i}|^2}},\quad 
|C^{'}_{R_{k}}|^2=\frac{|C_{R_{k}}|^2}{{1-\sum\limits_{i=0}^{k-2}|C_{R_i}|^2}}\nonumber.
\end{equation}
Non-zero components of the corresponding vector $\lambda^{(k-1)}$ are
\begin{equation}
\lambda^{(k-1)}_i=\begin{cases}
\frac{1}{p}|C^{'}_{R_{k-1}}|^2  &\text{if}\  i\in R_{k-1}\backslash R_k\cr
\frac{1}{p}|C^{'}_{R_k}|^2  &\text{if}\  i\in R_k\backslash R_{k-1}\cr
\frac{1}{p}\left [|C^{'}_{R_{k-1}}|^2+|C^{'}_{R_k}|^2 \right ]  &\text{if}\  i\in R_k\cap R_{k-1}\end{cases}\nonumber
\end{equation}
It can be easily shown that $\mu^{R_{k-1}}=b^{R_{k-1}}=\frac{1}{p}|C^{'}_{R_{k-1}}|^2$ and, consequently, 
$\lambda^{(k)}=v_{p\downarrow 1}(R_k)$ $\Box$
\begin{table}[hpbt]
\centering
\caption[]{Example of application of iteration formula (\ref{iter}) for the case $N=\{1,2,3,4\}$, $p=2$, and $\lambda^{(0)}=\left (\frac{7}{20},\frac{1}{4},\frac{1}{4},\frac{3}{20}\right )$ }
\begin{tabular}{lcccc}
\rule{0pt}{1pt}\\
\hline
\rule{0pt}{1pt}\\
$k$&$R_k$&Interval&$\mu^{R_k}$&$\lambda^{(k+1)}$\\
\rule{0pt}{1pt}\\
\hline
\rule{0pt}{15pt}
0  &$\{1,2\}$&$(0,\frac{1}{4}]$&$\frac{1}{5}$&$(\frac{1}{4},\frac{1}{12},\frac{5}{12},\frac{1}{4})$\\
\rule{0pt}{15pt}
1  &$\{1,3\}$&$(0,\frac{1}{4}]$&$\frac{1}{6}$&$(\frac{1}{8},\frac{1}{8},\frac{3}{8},\frac{3}{8})$\\
\rule{0pt}{15pt}
2  &$\{1,4\}$&$(0,\frac{1}{8}]$&$\frac{1}{16}$&$(\frac{1}{14},\frac{1}{7},\frac{3}{7},\frac{5}{14})$\\
\rule{0pt}{15pt}
3  &$\{2,3\}$&$(0,\frac{1}{7}]$&$\frac{1}{7}$&$(\frac{1}{10},0,\frac{2}{5},\frac{1}{2})$\\
\rule{0pt}{15pt}
4  &$\{2,4\}$&$ \{0\}$&$0$&\\
\rule{0pt}{15pt}
5  &$\{3,4\}$&$(0,\frac{2}{5}]$&$\frac{2}{5}$&$(\frac{1}{2},0,0,\frac{1}{2})$\\
\rule{0pt}{1pt}\\
\hline
\end{tabular}
\label{tab:1}
\end{table}
Now we may suggest the following scheme for generation of non-boundary expansions. First it is necessary to arrange $p-$element subsets from $Ind(\lambda^{(0)})$ in some fixed order $R_0,R_1,\ldots,R_{k_f}$ and step by step split vertices first  $v_{p\downarrow 1}(R_0)$ from $\lambda^{(0)}$, then $v_{p\downarrow 1}(R_1)$ from $\lambda^{(1)}$, etc, selecting at each step a non-boundary value of parameter $\mu^{R_i}$. This procedure is continued till the moment when on some step $k$ the number of remaining subsets $R_{k+1}, \ldots, R_{k_f}$ becomes equal to the number of active indices in the residual vector $\lambda^{(k)}$ minus 1. Then to this residual vector the algorithm of generation of the boundary expansion is applied. Simple example of an expansion, obtained in such a way, is given in Table \ref{tab:1} for the case $\lambda^{(0)}=\left (\frac{7}{20},\frac{1}{4},\frac{1}{4},\frac{3}{20}\right )$ and $p=2$. 

\bigbreak
\bigbreak

{\bf IV. Algebraic Version of The Hohenberg-Kohn Theorem}
\bigbreak
 
 The set of Hermitian linear operators over $p$-electron sector ${\cal F}_N^p$ of the Fock space will be denoted as  ${\EuScript H}_p$.

The mapping   
\begin{equation}
X:v\to \sum\limits_{i=1}^p\underbrace{I\otimes  \ldots \otimes I}_{i-1  }\otimes v\otimes I\ldots \otimes I
\label{X}
\end{equation}
defines a basis independent linear representation of the set of 1-electron operators ${\EuScript H}_1$ by $p-$electron operators from
${\EuScript H}_p$. 
Operators $X(v)$ are called `1-electron operators acting on $p$-electron space'. More simple but basis dependent definition of $X(v)$ may be given in terms of the standard creation-annihilation operators associated with some orthonormal MSO basis set.  

Let $H\in {\EuScript H}_p$ be some fixed operator over $p-$electron space and let
\begin{equation}
P_1^H(v)=\sum\limits_{j=1}^{\varkappa}|\Psi_1^{(j)}(v)><\Psi_1^{(j)}(v)|
\label{gsp}
\end{equation}
be the projector on the eigenspace $W_1^H(v)$ of operator $H+X(v)$ corresponding to its lowest eigenvalue $E_1^H(v)$. Let us introduce the following equivalence relation on the set ${\EuScript H}_1$:
\begin{equation}
R_H(v,v')\Leftrightarrow P_1^H(v)=P_1^H(v')
\label{eqrel}
\end{equation}
Quotient of ${\EuScript H}_1$ modulo the equivalence relation (\ref{eqrel}) contains classes of equivalent 1-electron operators and the notation ${\EuScript Q}_H={\EuScript H}_1/R_H$ for the  set of such classes will be used. Class represented by some 1-electron operator $v$ is denoted by the standard symbol $[v]$. 
The set of scalar operators is a subset of the zero class $[0]$ not necessarily coinciding with it.
   
If otherwise is not stated,  the quotient topology is supposed to be fixed on the set ${\EuScript Q}_H$. It is to be emphasize  that the definition of the set  ${\EuScript Q}_H$ in the  case of infinite dimension should be modified to exclude 1-electron operators $v$ such that the discrete part of spectrum of the operators $H+X(v)$ is empty.    
\begin{theorem}[Hohenberg-Kohn] The mapping
\begin{equation}
i_H:[v]\longrightarrow P_1^H([v])\longrightarrow \frac{1}{p!}c^{p-1}\left [\frac{1}{\varkappa}P_1^H([v])\right ]
\label{inj}
\end{equation}
is injective for any $H\in {\EuScript H}_p$.
\label{HK}
\end{theorem}
\textbf{Proof.} It is sufficient to prove that $[v]\ne [v']\Rightarrow i_H([v])\ne i_H([v'])$. Let us suppose the contrary, that is $[v]\ne [v']$ but $i_H([v])=i_H([v'])=\rho$.
Since, by definition, $[v]\ne [v']$ implies $W_1^H(v)\ne W_1^H(v')$, it is possible to select a pair of eigenfunctions $\Psi_1(v)\in W_1^H(v)$ and $\Psi_1(v')\in W_1^H(v')$ such that 
$\Psi_1(v)\not \in W_1^H(v')$ and $\Psi_1(v')\not \in W_1^H(v)$. Now, following the original arguments of Hohenberg and Kohn, we take into account that the quadratic functional associated with any Hermitian operator, reaches its absolute minimum at vectors belonging to the eigenspace of this operator corresponding to its lowest eigenvalue. We have two inequalities
\begin{multline*}
E_1^H(v)< Tr\left \{P_1^H(v')(H+X(v))\right \}=E_1^H(v')+ Tr \left \{P_1^H(v')\left [X(v)-X(v')\right ]\right \}\\
=E_1^H(v')+Tr\left \{\rho(v-v')\right\}\qquad\qquad\qquad\ 
\end{multline*}    
and
\begin{multline*}
E_1^H(v')< Tr\left \{P_1^H(v)(H+X(v'))\right \}=E_1^H(v)+ Tr \left \{P_1^H(v)\left [X(v')-X(v)\right ]\right \}
\} \\
=E_1^H(v)+Tr\left \{\rho(v'-v)\right \}\qquad\qquad\qquad\ 
\end{multline*}    
that contradict each other $\Box$

Let us point out the main differences between the original Hohenberg-Kohn statement \cite{Kohn-1,Kohn-2} and  theorem \ref{HK}:

(1) Theorem 4 states that the unique ensemble representable 1-density operator (not just density) corresponds to a certain class of 1-electron operators, but not vice versa;

(2) This theorem is valid for arbitrary choice of operator $H$ that can be, in particular, arbitrary 1-electron, or  even zero operator; 

(3) It is not presupposed that the mapping $i_H$ is continuous (with respect to the quotient topology on ${\EuScript Q}_H$ and the standard topology on  ${\EuScript V}_{N,p,1}$). 
 
Note as well that the statement  analogous to that in Theorem \ref{HK}  holds true for the set ${\EuScript H}_q$ of $q-$electron operators $(q<p)$   
with obvious replacement of $\frac{1}{p!}c^{p-1}\left [\frac{1}{\varkappa}P_1^H(v)\right ]$ in Eq.(\ref{inj}) by an ensemble representable density operator $\frac{q!}{p!}c^{p-q}\left [\frac{1}{\varkappa}P_1^H(v)\right ]$ of order $q$. This generalization, however, is of no particular interest because the structure of the set ${\EuScript V}_{N,p,q}$ of ensemble representable density operators of order $q$ is  unknown for $q>1$.  

It is easy to show that in general case the image of the mapping (\ref{inj}) does not coincide with the set ${\EuScript V}_{N,p,1}$ being only its proper subset. Indeed, let us suppose that the following  auspicious conditions hold true  (i) classes of  1-electron operators from  ${\EuScript H}_1$ can be continuously parametrized by free real parameters $t=(t_1,t_2,\ldots,t_k)\in \mathbb R^k$:    
\begin{equation}
[v(t)]=\left\{aI_n+v(t)\right \}_{a\in\mathbb R} 
\label{class}
\end{equation}     
($I_n$ is 1-electron identity operator), and (ii) the mapping $t\to P_1^H([v(t)])$ (see Eq.(\ref{gsp})) is continuous (and, consequently, the mapping (\ref{inj}) is also continuous). Even under such conditions 
the image of the mapping (\ref{inj}) is an open subset of the compact set ${\EuScript V}_{N,p,1}$.   It is pertinent to note as well that  the mapping (\ref{inj}) is not necessarily continuous  for arbitrary $H$. The simplest example corresponds to the case $H=0$ where the quotient set ${\EuScript Q}_H$ is discrete. 

In the finite dimensional case the operator space  ${\EuScript H}_1$ admits (in full analogy with the set of representable 1-density operators, see Section II) at least two convenient  realizations. In the first, standard,  realization the MSO orthonormal basis $\psi$ is supposed to be fixed and the operator space ${\EuScript H}_1$ coincides with the set of all unitary transformed operators that are diagonal in the basis under consideration. In the second realization a sum (disjoint union) of fibres isomorphic to the typical fibre $\mathbb R^n$ is considered:
\begin{equation}
{\widetilde {\EuScript H}}_1=\bigsqcup\limits_{\psi\in {\EuScript N}}{\EuScript H}_1^{\psi}
\end{equation}
where ${\EuScript H}_1^{\psi}$ is the set of 1-electron operators diagonal in the basis $\psi$. It is clear that there exists natural surjective mapping ${\widetilde {\EuScript H}}_1\to {\EuScript H}_1$. Of course, in the last realization the same 1-electron operator may appear on different fibres (for example, the 1-electron identity belongs to each fibre). This drawback is partially compensated by the relative simplicity of ${\widetilde {\EuScript H}}_1$ structure: it is homeomorphic to the Cartesian product the manifold ${\EuScript N}$ of all $n$-frames of 1-electron space (base) and the typical fibre $\mathbb R^n$. The set ${\widetilde {\EuScript H}}_1$ is called a total space of a (trivial) fibre bundle with the manifold ${\EuScript N}$ as its base, ${\EuScript H}_1^{\psi}$ as a fibre over point $\psi$, and $\mathbb R^n$ as the typical fibre.  

In analogy with 1-electron operators, for each $n$-frame $\psi$ operator $H$ has basis dependent realization $H_{\psi}$  with respect to the determinant basis $\{\psi_{i_1}\wedge\ldots\wedge\psi_{i_p}\}_{i_1<\ldots<i_p}$.  If $\psi '=\psi U$ then matrices of operators $H_{\psi}$ and $H_{\psi'}$ are connected by 
\begin{equation}
H_{\psi'}=\left (\bigwedge\limits^p U\right )^{\dagger}H_{\psi}\left (\bigwedge\limits^p U\right ) 
\end{equation}
Of course, $H_{\psi}$ and $H_{\psi'}$ have the same eigenspaces being different representations of the same operator. 

Since each fibre ${\EuScript H}_1^{\psi}$ is a subset of the set ${\EuScript H}_1$,  the equivalence relation (\ref{eqrel}) induces the equivalence relation  on ${\EuScript H}_1^{\psi}$ for each $n$-frame $\psi$. The corresponding equivalence classes  $[v]^{\psi}$ constitute the set ${\EuScript Q}_H^{\psi}$ that can be considered as a fibre of the fibre bundle
\begin{equation}   
{\widetilde {\EuScript Q}}_H=\bigsqcup\limits_{\psi\in {\EuScript N}}{\EuScript Q}_H^{\psi}
\end{equation}
The replacement of the set ${\EuScript Q}_H$ by the set ${\widetilde {\EuScript Q}}_H$ leads to the violation of the Hohenberg-Kohn theorem: the mapping 
\begin{equation}
{\tilde i_H}:[v]^{\psi}\longrightarrow P_1^H([v]^{\psi})\longrightarrow \frac{1}{p!}c^{p-1}\left [\frac{1}{\varkappa}P_1^H([v]^{\psi})\right ]
\label{finj}
\end{equation}
is obviously not injective.
However, this theorem  is valid for each fibre  ${\EuScript H}_1^{\psi}$, and even in more strong form: since  1-electron term $Tr[\rho v]=Tr[d_{\psi}(\rho)v]$ in average energy involves diagonal of representable 1-density operator, it is possible to prove that the mapping $d_{\psi}\circ {\tilde i_H}$ is injective for any $\psi$.   

Thus, we can formulate the algebraic version of the Hohenberg-Kohn statement as

\begin{theorem}
{\it The mapping $d\circ {\tilde i_H}=\{d_{\psi}\circ {\tilde i_H}\}_{\psi\in{\EuScript N}}$  is a fibrewise injective 
mapping of the fibre bundle ${\widetilde {\EuScript Q}}_H$ into the fibre bundle ${\widetilde {\EuScript V}}_{N,p,1}$. }
\end{theorem}
{\rm\bf Corollary.} {\it Restriction of ${\tilde i_H}$ on any fibre ${\EuScript Q}_H^{\psi}$ is injective for any $n$-frame $\psi$.}

Let us consider two in a certain sense boundary cases. 

For $H=0$ any fibre ${\EuScript Q}_0^{\psi}$ is a discrete topological space with a finite number of elements 
(classes $[v]^{\psi}$). Typical class $[v]_{k,l}^{\psi}$ includes infinitely many diagonal operators  
\begin{equation}
v=\sum\limits_{j=1}^n\mu_j|\psi_j><\psi_j|
\label{vres}
\end{equation}   
with the following restrictions on their eigenvalues
\begin{equation}
\max\limits_{j=1}^{p-k}\{\mu_j\}<\mu_{p-k+1}=\ldots =\mu_p=\ldots
=\mu_l< \min\limits_{j=l+1}^n\{\mu_j\}
\label{order}
\end{equation}
The ground state eigenspace of the corresponding operator $X(v), v\in [v]_{k,l}^{\psi}$ is of the dimension $\binom{l-p+k}k$. By means of permutations of indices $1,2,\ldots,n$ applied to the typical inequality (\ref{order}) it is easy to obtain $n!/[(p-k)!(l-p+k)!(n-l)!]$ classes associated with the selected typical class.  Elements of these classes are parametrized by $n+p-l-k+1$ real parameters of which $p-k$ parameters are free and the remainder ones obey inequality type restrictions.  The total number of classes is 
\begin{equation}
|{\EuScript Q}_0^{\psi}|=\binom{n}p+\sum\limits_{k=1}^p\binom{n}{p-k}\sum\limits_{l=p+1}^n\binom{n-p+k}{n-l}.
\end{equation}
For example, for $N=\{1,2,3,4\}$ and $p=2$ we have 

Six classes corresponding to single determinant ground states with typical inequality  $\max\{\mu_1,\mu_2\}<\min\{\mu_3,\mu_4\}$ and typical ground eigenspace $\mathbb C|12\rangle$; 

Twelve classes corresponding to 2-dimensional ground eigenspace with typical inequality  $\mu_1<\mu_2=\mu_3<\mu_4$ and typical ground eigenspace $\mathbb C|12\rangle\oplus\mathbb C|13\rangle$;    

Four classes corresponding to 3-dimensional ground eigenspace with typical inequality $\mu_1<\mu_2=\mu_3=\mu_4$ and typical ground eigenspace $\mathbb C|12\rangle\oplus\mathbb C|13\rangle\oplus\mathbb C|14\rangle$;

Four classes corresponding to 3-dimensional ground eigenspace with typical inequality $\mu_1=\mu_2=\mu_3<\mu_4$ and typical ground eigenspace $\mathbb C|12\rangle\oplus\mathbb C|13\rangle\oplus\mathbb C|23\rangle$;

One class of the identity operator with typical inequality $\mu_1=\mu_2=\mu_3=\mu_4$. 

The image ${\tilde i}_0([v]_{k,l}^{\psi})$ of the typical class $[v]_{k,l}^{\psi}$ is the density operator 
\begin{equation}
\rho=\frac{1}{p}\left [\sum\limits_{j=1}^{p-k}|\psi_j\rangle\langle\psi_j|+\frac{k}{k+l-p}\sum\limits_{j=p-k+1}^{l}|\psi_j\rangle\langle\psi_j|\right ].
\end{equation}

Now let us consider the case when $p$-electron operator $H$ defines the following structure of classes on the fibres ${\EuScript Q}_H^{\psi}$:
\begin{equation}
[v(t)]^{\psi}=\left \{ aI_n+\sum\limits_{j=1}^{n-1}t_j|\psi_j\rangle\langle\psi_j|\right \}_{a\in\mathbb R}
\label{classstr}
\end{equation}  
It is clear that this set of classes ${\EuScript Q}_H^{\psi}$ is homeomorphic to $\mathbb R^{n-1}$ and each class is homeomorphic to $\mathbb R$. If we suppose that the projector on the ground state of the operator $H+X(v(t))$ depends  continuously on parameters $t_1,t_2,\ldots,t_{n-1}$ then the image of the fibre ${\EuScript Q}_H^{\psi}$ is an open subset in ${\EuScript V}_{N,p,1}$ (or even in ${\EuScript P}_{N,p,1}$ if the lowest eigenvalue of the parametric operator $H+X(v(t))$ is non-degenerate for all $t$). This means that with the class structure defined by Eq.(\ref{classstr}) the image ${\tilde i}_H({\EuScript Q}_H^{\psi})$ can not contain 1-density operators from the border of ${\EuScript V}_{N,p,1}$ (or ${\EuScript P}_{N,p,1}$).  All aforementioned arguments are applicable only if the ground state projector of operator $H+X([v]^{\psi})$ is continuous (with respect to the quotient topology on  ${\EuScript Q}_H^{\psi}$). And  between these two boundary cases there exists a variety of  'intermediate' ones with more complicated structure of sets of classes.    

Parametrization of the set of classes defined by Eq.(\ref{classstr}) is by no means unique. Sometimes it is convenient, for example, to impose the additional requirement of orthogonality of the parametric representative of class $[v(t)]^{\psi}$ to the identity operator (say, with respect to the trace scalar product). It may be easily done by 
writing class $[v(t)]^{\psi}$ in the form                 
\begin{equation}
[v(t)]^{\psi}=\left \{ aI_n+\sum\limits_{j=1}^nt_j|\psi_j\rangle\langle\psi_j|\right \}_{a\in\mathbb R}
\label{classstr1}
\end{equation}  
with the additional condition $\sum\limits_{j\in N}t_j=0$. Concrete examples of different parametrizations will be given later.   

Now let us turn to more detailed analysis of the 'diagonal' version of the Hohenberg-Kohn statement  and consider the mapping $d_{\psi}\circ {\tilde i_H}: {\EuScript Q}_H^{\psi}\to {\sf V}^{\psi}_{N,p,1}$ for some fixed $n$-frame $\psi$. It will be supposed that matrix $H_{\psi}$ of $p$-electron operator $H$ has non-degenerate ground state. Parametric  matrix $H_{\psi}+X\left (v^{\psi}(t)\right )$ either has non-degenerate ground state for all finite values of parameters, or there are points in the parameter space where the ground state is degenerate, or quasi-degenerate. In strictly degenerate case, due to Theorem 5, the mapping ${\tilde i_H}$ should have discontinuity when going from pure to averaged 1-density diagonal. 

It should be specially emphasized that the structure of classes from ${\EuScript Q}_H^{\psi}$ depends on the properties of representable diagonal $d_{\psi}(\rho) $ of 1-density operator $\rho$ corresponding to the ground state of operator $H$. In particular, this structure is different for $d_{\psi}(\rho)\in {\overset{\circ}{\sf V}}{\vphantom V}^{\psi}_{N,p,1}$ and $d_{\psi}(\rho)\in \partial {\sf V}^{\psi}_{N,p,1}$. Indeed, let 
\begin{equation}
\rho_{ii}=
\begin{cases}
0&\mbox{if}\  i\in I\\
\frac{1}{p}&\mbox{if}\  i\in J
\end{cases}
\label{bound}
\end{equation}
where 
$I\cap J=\emptyset$. This means (see Appendix A) that  $d_{\psi}(\rho) $ belongs to the $(I,J)$-face of the polyhedron ${\sf V}^{\psi}_{N,p,1}$ of dimension $n-|I|-|J|-1$. Ground state wave function $\Psi_1$ of matrix $H_{\psi}$ corresponds of 1-density operator with diagonal elements satisfying conditions (\ref{bound}) if it is orthogonal to subspace spanned by  $p$-electron basis determinants $|R\rangle $ such that either $ R\cap I\ne \emptyset$ or $R\cap J=\emptyset$, and arbitrary matrix of the form
\begin{equation}
{\bar  H}_{\psi}(a,b)=H_{\psi}+aI_n+\sum\limits_{k=0}^{{\min\{|I|,p\}}}\sum\limits_{K\subset I}^{(k)}\sum\limits_{L\subset N\backslash  (I\cup J)}^{(p-k)}b_{K,L}|K\cup L\rangle\langle K\cup L| 
\label{param}
\end{equation} 
has $\Psi_1$ as its eigenvector. Here $I_n$ is 1-electron identity matrix. To describe the structure of the set ${\EuScript Q}_H^{\psi}$, it is necessary to find out under what  restrictions on the coefficients $b_{K,L}$ 

(i) the parametric term on the right hand side of Eq.(\ref{param}) is 1-electron matrix of the form of Eq.(\ref{X}), 

and 

(ii) ${\bar  H}_{\psi}$ has $\Psi_1$ as its lowest eigenvector.      
  
Let us consider  an example  formally corresponding to 2-electron case: $N=\{1,2,3,4\}$ and $p=2$. 

For a fixed 4-frame $\psi=(\psi_1,\psi_2,\psi_3,\psi_4)$ general two-electron wave function is of the form
\begin{equation}
\Psi =a\psi_1\wedge \psi_2+b\psi_1\wedge \psi_3+c\psi_1\wedge \psi_4+d\psi_2\wedge \psi_3+e\psi_2\wedge \psi_4+f\psi_3\wedge \psi_4
\label{examwf}
\end{equation}
Diagonal of 1-density operator $\rho$ corresponding to this wave function is
\begin{equation}
d_{\psi}(\rho)=\frac{1}{2}(a^2 + b^2 + c^2, a^2 + d^2 + e^2, b^2 + d^2 + f^2, c^2 + e^2 + f^2)
\end{equation}
We analyze three cases: (1) $d_{\psi}(\rho)$ belongs to the interior part of the polyhedron ${\sf V}^{\psi}_{N,2,1}$;
(2) $d_{\psi}(\rho)$ belongs to a 2-face of this polyhedron, and (3) $d_{\psi}(\rho)$ belongs to an edge of the polyhedron ${\sf V}^{\psi}_{N,2,1}$.   

With the aid of MATHEMATICA 4.1 package \cite{math} three $6\times 6$-matrices $H_{\psi}^{(1)}$, $H_{\psi}^{(2)}$, and  $H_{\psi}^{(3)}$ with predefined spectrum and eigenfunctions, corresponding to these  three cases, were generated (see Appendix C).   
 
General form of 1-electron operators belonging to the fibre ${\EuScript H}_1^{\psi}$ is  
\begin{equation}
v^{\psi}(t)=
\begin{pmatrix}
t_1&0&0&0\\
0&t_2&0&0\\
0&0&t_3&0\\
0&0&0&t_4
\end{pmatrix}
\label{vt}
\end{equation}
and the corresponding 2-electron operator may be written as (see Eq.(\ref{X})
\begin{equation}
X\left (v^{\psi}(t)\right )=
\begin{pmatrix}
t_1+t_2&0&0&0&0&0\\
0&t_1+t_3&0&0&0&0\\
0&0&t_1+t_4&0&0&0\\
0&0&0&t_2+t_3&0&0\\
0&0&0&0&t_2+t_4&0\\
0&0&0&0&0&t_3+t_4
\end{pmatrix}
\label{Xv0}
\end{equation}
 
{\bf Case 1.} $d_{\psi}(\rho)=(0.404545, 0.313636, 0.163636, 0.118182)\in {\overset{\circ}{\sf V}}{\vphantom V}^{\psi}_{N,2,1}$.

The lowest eigenfunction $\Psi_1^{(1)}$ of the matrix $H_{\psi}^{(1)}$ involves all 6 determinants and all coefficient $b_{K,L}$ in Eq.(\ref{param}) should be taken equal to zero. 
This is the most general case and classes of 1-electron operators are of the form
\begin{equation}
[v(x,y,z)]^{\psi}=\left \{aI_4+\begin{pmatrix}
x&0&0&0\\
0&y&0&0\\
0&0&z&0\\
0&0&0&-x-y-z
\end{pmatrix}\right \}_{a\in \mathbb R}
\end{equation}
where $x = 3 t_1 - t_2 - t_3 - t_4$, $y = -t_1 + 3 t_2 - t_3 - t_4$, and $z = -t_1 - t_2 + 3 t_3 - t_4$.
Each class is homeomorphic to $\mathbb R$ and the set of classes ${\EuScript Q}_H^{\psi}$ is homeomorphic to $\mathbb R^3$. Note that in this concrete case the set of classes carries natural vector space structure. 

\begin{figure}[ht]
	\centering
		\includegraphics*[width=0.70\textwidth]{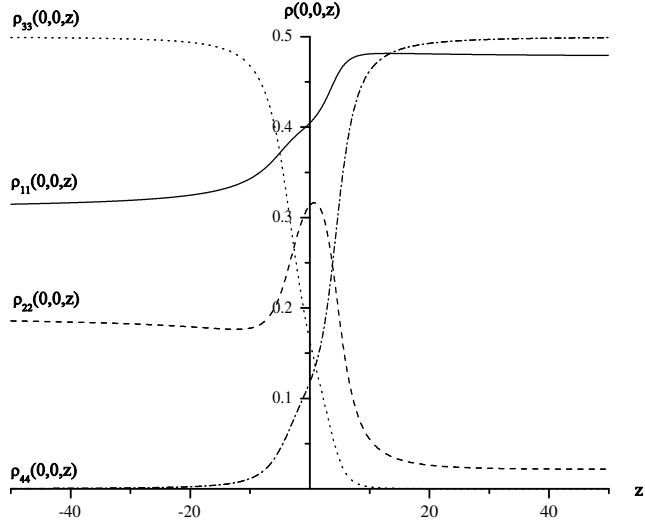}
	\caption{Diagonal matrix elements of parametric 1-density matrix $\rho(x,y,z)$: functions $z\to \rho_{ii}(0,0,z)$}
	\label{fig:pz}
\end{figure}

An example of behavior of diagonal elements of 1-density operators along one of  basis directions in the parameter space is displayed on Figs.\ref{fig:pz}. 
It is seen that these elements are smooth functions along the selected  direction and that the mapping $z\to (\rho_{11}(0,0,z),\rho_{22}(0,0,z),\rho_{33}(0,0,z),\rho_{44}(0,0,z))$ is injective. It may be expected that each representable diagonal corresponds to the unique set of parameters, and, consequently, to the unique class of 1-electron operators. It is also clear that for any finite values of parameters the corresponding diagonals of 1-density operators belong to the interior part of the polyhedron ${\sf V}^{\psi}_{N,2,1}$ in full accordance with the standard topological arguments. It seems to be most likely that for the operator under consideration
 the (topological) closure of the image of the set  ${\EuScript Q}_H^{\psi}$ with respect to the mapping $d_{\psi}\circ {\tilde i_H}$ coincides with the polyhedron ${\sf V}^{\psi}_{N,2,1}$. 

{\bf Case 2.} $d_{\psi}(\rho)=(0.333333,0.333333,0.333333,0)\in F_{\{4\},\emptyset}$ where $F_{\{4\},\emptyset}=Conv(v_{2\downarrow 1}(12),v_{2\downarrow 1}(13),v_{2\downarrow 1}(23))$ is 2-face
of the polyhedron  ${\sf V}_{N,2,1}^{\psi}$ (see Appendix A). 

\begin{figure}[ht]
	\centering
		\includegraphics*[width=0.70\textwidth]{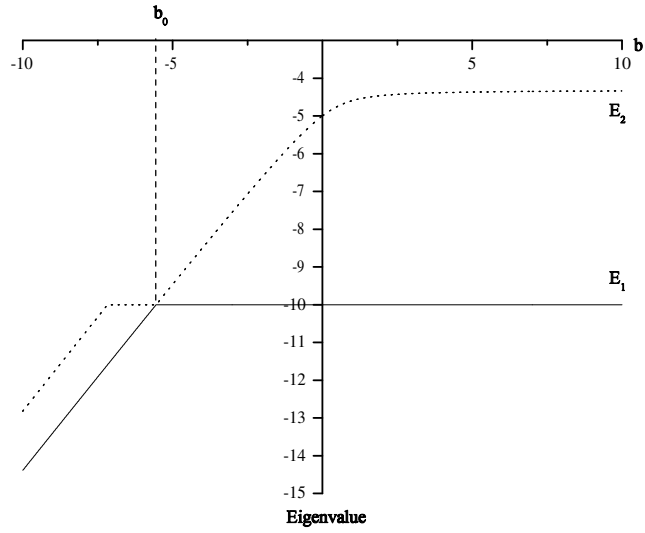}
	\caption{Two lowest eigenvalues of the parametric operator ${\bar  H}_{\psi}^{(2)}(a,b)$ (see Eq.(\ref{paramex})) as functions of parameter b}
	\label{fig:e}
\end{figure}

The lowest eigenvector $\Psi_1^{(2)}$ of matrix $H_{\psi}^{(2)}$ is orthogonal to basis determinants $|14\rangle ,|24\rangle$, and $|34\rangle$. However, the parametric operator (see Eq.(\ref{param}))
\begin{equation}
{\bar  H}_{\psi}^{(2)}(a,b)=H_{\psi}^{(2)}+\frac{1}{2}aX(I_4)+bX(J_4)
\label{paramex}
\end{equation} 
where  
\begin{equation}
J_4=\begin{pmatrix}
0&0&0&0\\
0&0&0&0\\
0&0&0&0\\
0&0&0&1
\end{pmatrix}
\end{equation}
has $\Psi_1^{(2)}$ as its lowest eigenvector only for $b_0\le b < \infty $ where $b_0\approx -5.5484$ (see Fig.\ref{fig:e} ). 
As a result, at the point $b_0$ the class structure undergoes essential change. We have 
\begin{equation}
[v]^{\psi}=\begin{cases}
\left \{aI_4+bJ_4+\begin{pmatrix}
x&0&0&0\\
0&y&0&0\\
0&0&-x-y&0\\
0&0&0&0
\end{pmatrix}\right \}_{(a,b)\in \mathbb R\times (b_0,\infty)} &\mbox{if}\  b>b_0\\
\left \{aI_4+\begin{pmatrix}
x&0&0&0\\
0&y&0&0\\
0&0&-x-y-b&0\\
0&0&0&b
\end{pmatrix}\right \}_{a\in \mathbb R} &\mbox{if}\  b\le b_0
\end{cases}
\end{equation}

{\bf Case 3.} $d_{\psi}(\rho)=(0.5, 0.470588,0.0294118,0)\in F_{\{4\},\{1\}}$ where $F_{\{4\},\{1\}}=Conv(v_{2\downarrow 1}(12),v_{2\downarrow 1}(13))$ is 1-face (edge) of the polyhedron  ${\sf V}_{N,2,1}^{\psi}$ (see Appendix A).

\begin{figure}[ht]
	\centering
		\includegraphics*[width=0.70\textwidth]{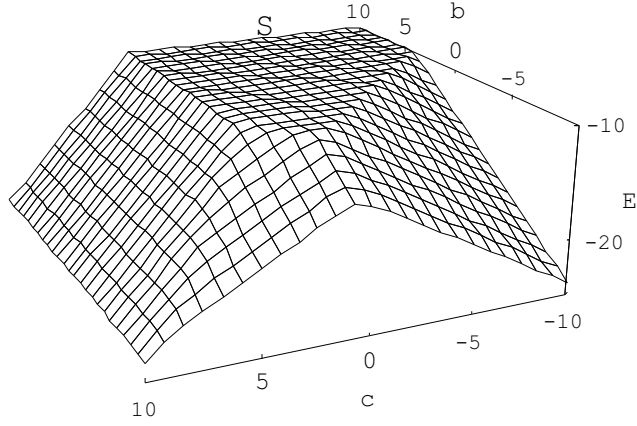}
	\caption{The lowest eigenvalue of the parametric operator ${\bar  H}_{\psi}^{(3)}(a,b,c)$ (see Eq.(\ref{param-3})) as a function of parameters $b$ and $c$}
	\label{fig:ebc}
\end{figure}

The lowest eigenvector $\Psi_1^{(3)}$ of matrix $H_{\psi}^{(3)}$ is orthogonal to basis determinants $|14\rangle$, $|23\rangle$, $|24\rangle$, and $|34\rangle$. The auxiliary parametric operator 
\begin{equation}
\bar H_{\psi}^{(3)}(a,b,c)=H_{\psi}^{(3)}+\frac{1}{2}aX(I_4)+bX(J_4)+cX(K_4)
\label{param-3}
\end{equation}
where
\begin{equation}
K_4=\begin{pmatrix}
1&0&0&0\\
0&-1&0&0\\
0&0&-1&0\\
0&0&0&0
\end{pmatrix}
\end{equation}
has $\Psi_1^{(3)}$ as its eigenfunction. To determine the relevant class structure, it is necessary to find out when the parametric operator (\ref{param-3}) has $\Psi_1^{(3)}$ as its lowest eigenvector.  From Fig.\ref{fig:ebc} it is seen that the set of admissible parameters is a certain subset $S$ of the plane $E=10$ in 3-dimensional space of triples $(b,c,E)$. 
Within this subset the class structure is 
\begin{equation}
[v(x)]^{\psi}=
\left \{aI_4+bJ_4+cK_4+\begin{pmatrix}
0&0&0&0\\
0&x&0&0\\
0&0&-x&0\\
0&0&0&0
\end{pmatrix}
\right \}_{(a,b,c)\in\mathbb R\times  S}
\end{equation}
After passing through the border of the set $S$ the class structure changes. 

Thus, sets ${\EuScript Q}_H^{\psi}$ of classes of 1-electron operators associated with some $p$-electron operator $H_{\psi}$ may be of a rather complicated nature. It may be a union of cells of different dimensions as it was in the last two examples, and  use of general parametric classes of the type of Eq.(\ref{classstr1}) for both interior and border points of the polyhedron ${\sf V}^{\psi}_{N,p,1}$ will lead to violation of Theorem 5. Probably the only regular case corresponds to the situation when representable diagonal of 1-density operator, corresponding to the ground state of operator $H$, belongs to the interior part the Coleman's polyhedrons ${\sf V}_{N,p,1}^{\psi}$ for any $n$-frame $\psi$ (see Proposition 2). In this case the following condition  may be fulfilled:  
\begin{equation}
d_{\psi}\circ {\tilde i_H}\left ({\EuScript Q}_H^{\psi}\right )={\overset{\circ}{\sf V}}{\vphantom V}^{\psi}_{N,p,1}    \label{HR}
\end{equation}
for any $n$-frame $\psi$. If the equalities (\ref{HR}) hold true, then the inverse  mappings 
\begin{equation}
[d_{\psi}\circ {\tilde i_H}]^{-1}: {\overset{\circ}{\sf V}}{\vphantom V}^{\psi}_{N,p,1}\to {\EuScript Q}_H^{\psi} 
\label{inverse}
\end{equation}
are correctly defined.
 
Till now we studied in a certain sense local task of parametrization of representable diagonals by classes of diagonal 1-electron operators from a given fibre ${\EuScript H}_1^{\psi}$. The next very important step is to assemble  local results.  Since the fibre bundles under consideration are trivial (that is not twisted, in contrast to, say,  the famous M\"obius band), this task, in theory,  is not complicated. Choosing any convenient (local) parametrization of the base $\EuScript N$ ($\dim \EuScript N=n(n-1)/2$ if the ground number field is $\mathbb R$),  we come to  1-density diagonals depending on two independent sets of parameters. The first set parametrizes some open subset of the base $\EuScript N$ and the second set parametrizes classes of diagonal 1-electron operators. 

Now let us consider the case when for each $n$-frame $\psi$ the structure of classes of 1-electron operators from    
${\EuScript H}_1^{\psi}$ is given by Eq.(\ref{classstr1}) without explicit reference to any concrete $p$-electron operator $H$. Each fibre ${\EuScript Q}^{\psi}$ (subscript $H$ is omitted because of the aforementioned reason)  is $(n-1)$-dimensional vector space over $\mathbb R$ with respect to the operations   
\begin{equation}
[v(t)]^{\psi}+[v(t')]^{\psi}=[v(t+t')]^{\psi},\quad  \alpha [v(t)]^{\psi}=[v(\alpha t)]^{\psi}, \quad \alpha\in \mathbb R
\end{equation}
and the corresponding fibre bundle $\widetilde {\EuScript Q}$ together with the vector space structure on each fibre is  the so-called vector bundle \cite{Wells}. The following problem seems to be of primary interest for the DFT theory:

Under what conditions  the inverse of a fibrewise bijective mapping
\begin{equation}
s:\bigsqcup\limits_{\psi\in {\EuScript N}}{\overset{\circ}{\sf V}}{\vphantom V}^{\psi}_{N,p,1}\to \bigsqcup\limits_{\psi\in {\EuScript N}}{\EuScript Q}^{\psi}
\label{inverse1}
\end{equation}
is the Hohenberg-Kohn mapping $d\circ {\tilde i_H}$ for some $p$-electron operator $H$?

It is easy to present  the following very important necessary condition that should be imposed on a fibrewise mappings of the type of Eq.(\ref{inverse1}): 

For any fixed $n$-frame $\psi$ and any 1-electron unitary transformation $U$
\begin{equation}
s_{\psi U}\left (d_{\psi U}\left ( U^{\dagger}\rho_{\psi}U\right )\right )=[0]^{\psi U}
\end{equation}    
where $\rho_{\psi}$ is some 1-density operator belonging to the interior part of the set ${\EuScript V}_{N,p,1}$.

In concluding this section let us discuss in general form the universal functionals of the DFT theory (see \cite{Yang, Kohn-2} and references therein). Relevant definitions and theorems from set-theoretical topology may be found in \cite{Bourbaki}. 
\begin{figure}[ht]
	\centering
		\includegraphics*[width=0.70\textwidth]{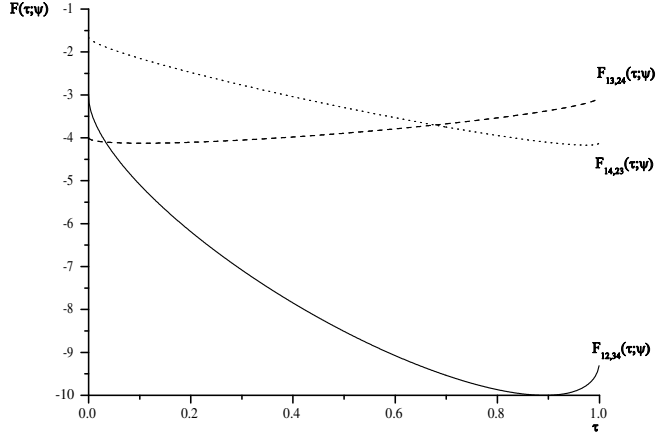}
	\caption{Branches of the universal functional on a fibre ${\sf P}^{\psi}_{N,2,1}$ corresponding to the  natural MSO 4-frame of operator ${H}_{\psi}^{(1)}$ ground state }
	\label{fig:univ}
\end{figure}
Let us suppose that 
\begin{equation}
\gamma:\mathbb P({\cal F}_{N,p})\to {\EuScript T}
\end{equation}
is a continuous surjective mapping of the (projective) space of states on a Hausdorff topological space  ${\EuScript T}$. Being the image of the compact set $\mathbb P({\cal F}_{N,p})$ with respect to continuous mapping, ${\EuScript T}$  is also compact. Let $R_{\gamma}$ be the equivalence relation associated with the mapping $\gamma$:
\begin{equation}
R_{\gamma}(x,y)\Leftrightarrow \gamma(x)=\gamma(y)
\end{equation}   
The quotient $\mathbb P({\cal F}_{N,p})/R_{\gamma}$ is necessarily  Hausdorff and compact space (in quotient topology) and, consequently, the mapping $\gamma$ admits representation in the form $\gamma=g\circ \pi$ where $\pi$ is the canonical projection $\mathbb P({\cal F}_{N,p})\to\mathbb P({\cal F}_{N,p})/R_{\gamma}$ and $g$ is the homeomorphism $\mathbb P({\cal F}_{N,p})/R_{\gamma}\to {\EuScript T}$.  The average energy $E_H=\langle \Psi|H|\Psi\rangle$ is a smooth mapping $\mathbb P({\cal F}_{N,p})\to \mathbb R$. Of course, this mapping is not compatible with the equivalence relation $R_{\gamma}$. But it is possible to introduce a new mapping ${\widetilde E}_H$ as
\begin{equation}
{\widetilde E}_H:\gamma^{-1}(\tau)\to \inf\limits_{x\in \gamma^{-1}(\tau)}E_H(x),\quad \tau\in {\EuScript T}
\end{equation} 
that is correctly defined due to the Weierstrass theorem \cite{Bourbaki}. The universal functional may be written as $F={\widetilde E}_H\circ g^{-1}:{\EuScript T}\to \mathbb R$. It is continuous if the mapping ${\widetilde E}_H$ is continuous. 

If $\gamma=\frac{1}{p!}c^{p-1}$ then  ${\EuScript T}={\EuScript P}_{N,p,1}$.  The domain of the universal functional $F:{\EuScript P}_{N,p,1}\to\mathbb R$, even if the mapping ${\widetilde E}_H$ is continuous, not necessarily admit global parametrization (see Appendix B). Therefore,  formal manipulations involving derivatives and variations of universal functionals, require certain care.

Let us suppose that the set   ${\EuScript P}_{N,p,1}$ is realized as a fibre bundle (\ref{pure}) and consider simple example: $N=\{1,2,3,4\}$ and $p=2$ (see Appendix B). In this case there are three branches of universal functional associated with three pairs of disjoint subsets: $F_{R,N\backslash R}(\lambda;\psi)$.
If 2-electron operator $H_{\psi}^{(1)}$  (see Appendix C) is chosen, then the ground state branch corresponds to the pair $(12,34)$ (it is sufficient to find the ground state 1-density operator and then perform transformation to the natural MSO basis to see that in this basis $\Psi_1^{(1)}=0.944996|12\rangle - 0.327082|34\rangle$). Dependence of 3 branches of the the universal functional on the parameter $\tau$ (see Appendix B, Eq.(B.2)) on the fibre ${\sf P}^{\psi}_{N,2,1}$, corresponding to the ground state natural 4-frame $\psi$, is displayed on Fig. \ref{fig:univ}. In the case  under consideration these branches are just segments of three ellipses.     

\bigbreak
{\bf V. Conclusion }
\bigbreak

Any finite dimensional model of quantum mechanics is in a certain sense incomplete, because it does not include many important relations like commutation relations of canonically conjugate pairs of observables which exist only in infinite dimension. On the other hand,   in the majority of cases molecular calculations are performed in finite basis sets, and it seems reasonable to try to  find out what specific finite dimensional features are  inherent in algebraic version of quantum chemistry methods. Thorough analysis of the algebraic version of the Hohenberg-Kohn theorem for arbitrary selected many electron operator $H$ shows that 

 - structure of classes of 1-electron operators that appear in the Hohenberg-Kohn theorem may be very complicated and may be determined not only by operator $H$ ground state but also by spectrum of a certain auxiliary parametric many electron operator; this structure strongly depends on the properties of the operator $H$ ground state 1-density operator $\rho$ and is different for $\rho$ belonging to the border of the Coleman's set and to its interior part;  

 - if $\rho$ belongs to the interior part of the Coleman's set (all MSO's are necessarily active) then there exist many electron operators such that the corresponding classes of 1-electron operators are of general form $\left \{aI_n+v(t)\right \}_{a\in \mathbb R}$; in this case  the Hohenberg-Kohn mapping may parametrize the interior part of the Coleman's set  but its image can not contain, say, vertices of this set (HF 1-density operators);   

 - the Hohenberg-Kohn mapping is not necessarily continuous for arbitrary operator $H$;

 - in finite dimensional case fundamental role is played not by densities but by representable diagonals of 1-density operators;
 
 - the set of 1-density operators representable by pure states, in general, does not admit global parametrization and there may exist several branches of DFT universal functionals. 

General theory developed in this work does not impose any special restrictions on many electron operator $H$. However, in all examples considered it was assumed that the ground state of this operator is non-degenerate. Degenerate case      may introduce additional complications in concrete structure of classes of 1-electron operators and  requires separate investigation.

\bigbreak

{\bf ACKNOWLEDGMENTS}
 
 The author  gratefully acknowledges the Russian Foundation for
Basic Research (Grant 06-03-33060) for financial support of the
present work.

\bigbreak
{\bf Appendix A: Combinatorial Structure of Polyhedron ${\sf V}_{N,p,1}$}
\bigbreak

Polyhedron ${\sf V}_{N,p,1}$ is situated in the (affine) hyperplane 
$$
{\sf H}_{\mathfrak a}=\{x\in\mathbb R^n:\langle \mathfrak a|x\rangle =1\}
\eqno(A.1)
$$
of the vector space $\mathbb R^n$ that has the vector
$$
\mathfrak a=\sum\limits_{i=1}^n e_i
\eqno(A.2)
$$
as its normal. Here $\{e_i\}_{i\in N}$ is the canonical basis of $\mathbb R^n$ and $N=\{1,2,\ldots,n\}$.

It is convenient to recast the Coleman's system (10) in a form 
commonly accepted in theory of polyhedral sets:
$$
\begin{cases}
\langle e_i|\lambda\rangle\ge 0,\ i\in N\\
\langle \mathfrak a-p e_i|\lambda\rangle\ge 0,\ i\in N\\
\langle \mathfrak a|\lambda\rangle =1
\end{cases}
\eqno(A.3)
$$
To describe faces of the polyhedron ${\sf V}_{N,p,1}$ it is necessary to analyze  
'mixed' families $\{\{e_i\}_{i\in I},\{\mathfrak a-pe_j\}_{j\in J},\mathfrak a\}$ with $I\cap J=\emptyset$. Two obvious restrictions should be imposed on disjoint subsets $I$ and $J$: $0\le |I|\le n-p$ and $0\le |J|\le p$. Indeed, if $|I|>n-p$ then the number of non-vanishing components of vector $\lambda\in {\sf V}_{N,p,1}$ is less than $p$ and consequently, $\langle \mathfrak a|\lambda\rangle <1$. 
If $|J|>p$ then the number of components of vector $\lambda\in {\sf V}_{N,p,1}$ equal to $\frac{1}{p}$ is greater than $p$ and, consequently, $\langle \mathfrak a|\lambda\rangle >1$. Direct calculation shows  that a family $\{\{e_i\}_{i\in I},\{\mathfrak a-pe_j\}_{j\in J},\mathfrak a\}$ is free if and only if $I\cap J =\emptyset$ and $|I\cup J|<n$. Thus, faces of the polyhedron ${\sf V}_{N,p,1}$ are the sets $F_{I,J}$ of solutions of the systems:
$$
\begin{cases}
\langle e_i|\lambda\rangle = 0,\ i\in I\\
\langle \mathfrak a-p e_j|\lambda\rangle= 0,\ j\in J\\
\langle \mathfrak a|\lambda \rangle =1\\
\langle e_i|\lambda\rangle \ge 0,\ i\in N\backslash I\\
\langle \mathfrak a-p e_j|\lambda\rangle\ge 0,\ j\in N\backslash J
\end{cases}
\eqno(A.4)
$$

Let $I, J$ be subsets of the index set $N$, such that $I\cap J=\emptyset $,  $|I\cup J|\le n-1$, $0\le |I|\le n-p$,  $0\le|J|\le p$, and let $K=N\backslash (I\cup J)$. Let us put $|I\cup J|=r$ and $|J|=p-s$. Admissible values of $r$ are $1,2,\ldots,n-1$. For fixed value of $r$ the admissible values of $|J|$ are $\min\{r,p\},\min\{r,p\}-1,\ldots,\max\{0,r-n+p\}$. If $\lambda  \in F_{I,J}$ then the following equality holds true 
$$\frac{|J|}{p}+\sum\limits_{k\in K}\lambda_k =1
\eqno(A.5)
$$
This equality can be recast in a more convenient form:
$$\sum\limits_{k\in K}\lambda_k=\frac{s}{p}
\eqno(A.6)
$$
where $s=\max\{p-r,0\},\max\{p-r,0\}+1,\ldots,\min\{p,n-r\}$. There are three cases to be analyzed.

(1) $s=\max\{p-r,0\}=0$.

In this case Eq.(A.6) has only trivial solution $\lambda_k=0,\  k\in K$. The corresponding face $F_{I,J}$ is just the vertex $v_{p\downarrow 1}(J)$ (0-face).

(2) $s\ne 0$ and $|K|\ge s+1$. 

The set of solutions of Eq.(A.6) is non-degenerate Coleman's polyhedron $\frac{s}{p}{\sf V}_{K,s,1}$ of the dimension $|K|-1$ and, consequently, $(|K|-1)$-face of the polyhedron ${\sf V}_{N,p,1}$. 

(3) $s\ne 0$ and $|K|=s=n-r$.

In this case we have degenerate Coleman's polyhedron $\frac{s}{p}{\sf V}_{K,s,1}=\{\frac{s}{p}v_{s\downarrow 1}(K)\}$ of dimension 0 (vertex) and, consequently, 0-face $v_{p\downarrow 1}(J\cup K)$ of  the polyhedron ${\sf V}_{N,p,1}$.

Now we can calculate the number of faces of a given dimension $n-r-1$ of the polyhedron ${\sf V}_{N,p,1}$:
$$
f_{n-r-1}=
\begin{cases} \binom{n}p &\mbox{if}\quad r=n-1\\
\sum\limits_{s=\max\{p-r,1\}}^{\min\{p,n-r-1\}}\binom{n}{p-s}\binom{n-p+s}{r-p+s} &\mbox{if}\quad r<n-1
\end{cases}
\eqno(A.7)
$$
For example, for $n=12$ and $p=5$ the $f$-vector (see, e.g. \cite{bron}) is $f=(1,792, 13860, 46200, 76230, 77616, 52668, 24552, 7920, 1760, 264, 24, 1)$ where we have added two coordinates $f_{-1}=1$ and $f_{n-1}=1$ corresponding to improper faces $\emptyset$ and ${\sf V}_{N,5,1}$. It is easy to check that this $f$-vector satisfies the classic Euler identity:
$$
\sum\limits_{k=0}^{n-1}(-1)^k f_k=0
\eqno(A.8)
$$  
Summing up, we can state that each $(n-r-1)$-face of dimension greater than 0 of the polyhedron ${\sf V}_{N,p,1}$  is defined by a pair of disjoint subsets $(I,J)$ with $|I\cup J|=r$ excluding the cases $|J|=p$ and $|J|=r-n+p$. A vertex $v_{p\downarrow 1}(R)$ belongs to this face if and only if $J\subset R\subset N\backslash I$. The number of vertices belonging to such a face is equal to $\binom{n-r}{p-|J|}$. Coleman's polyhedrons form a chain
$$
\left \{\frac{1}{n}\mathfrak a\right \}={\sf V}_{N,n,1}\subset {\sf V}_{N,n-1,1}\subset \ldots {\sf V}_{N,2,1}\subset {\sf V}_{N,1,1}
\eqno(A.9)
$$
where ${\sf V}_{N,1,1}$ and ${\sf V}_{N,n-1,1}$ are  standard and non-standard  simplexes of the vector space $\mathbb R^n$, respectively.  

\bigbreak
{\bf Appendix B: An Example of Geometric Description of the Set of 1-Density Operators Representable by Pure States}
\bigbreak

Let us consider  general two-electron wave function of the form of Eq.(\ref{examwf})
The corresponding pure state $|\Psi\rangle\langle\Psi|$ belongs to the projective space ${\mathbb P}({\cal F}_{N,2})$. We confine our consideration to the case when the ground number field is $\mathbb R$, and the projective space can be realized as a quotient of the 5-dimensional unit sphere ${\sf S}^5$ formed by pairs of diametrically opposite points. 

Matrix of 1-density operator with respect to fixed basis $\psi$ obtained by contraction of $|\Psi\rangle\langle\Psi|$ is 
$$
\rho=\frac{1}{2}
\begin{pmatrix}
a^2 + b^2 + c^2&bd + ce&-ad + cf&-ae - bf\\
      bd + ce&a^2 + d^2 + e^2&ab + ef&ac - df\\
      -ad + cf&ab + ef&b^2 + d^2 + f^2&bc + de\\
      -ae - bf&ac - df&bc + de&c^2 + e^2 + f^2
\end{pmatrix}\eqno(B.1)
$$  
Let us suppose that 1-density matrix (B.1) is diagonal in the basis under consideration. We have the system of 6 polynomial equations with respect to variables $a,b,c,d,e,f$ that can be easily solved with the aid of MATHEMATICA built-in {\bf Reduce} function \cite{math}. For each pair of 2-electron determinants with disjoint index sets there is a solution 
$$
\rho(R,N\backslash R)=\tau v_{2\downarrow 1}(R)+(1-\tau)v_{2\downarrow 1}( N\backslash R)
\eqno(B.2)
$$
where $R$ is 2-element subset of the index set $N$.

Using iteration formula (\ref{iter}) it is possible to construct infinitely many (for $\tau\ne 0,1$) 2-electron determinant ensembles contracted in diagonal (B.2). The corresponding pure states (see Proposition 1), however, give diagonal 1-density operators (over $\mathbb R$) only if  
$$
\Psi_{\pm}= \tau^{\frac{1}{2}}|R\rangle \pm (1-\tau)^{\frac{1}{2}}|N\backslash R\rangle
\eqno(B.3)
$$  
Thus, in the realization of the set ${\EuScript P}_{N,2,1}$ as a fibre bundle with the manifold of 4-frames as its base the fibre ${\sf P}^{\psi}_{N,2,1}$ over point $\psi$ consists of 1-density operators of the type of Eq.(B.2).
The  number of different types of such operators is equal to 3. In geometric terms the fibre ${\sf P}^{\psi}_{N,2,1}$ is a union of 3 closed line segments  having exactly one central point $(\frac{1}{4},\frac{1}{4},\frac{1}{4},\frac{1}{4})$ in common. Representable by pure 2-electron states 1-density operators, diagonal with respect to the basis $\psi$, necessarily have the form of Eq.(B.2), and, consequently, $\dim {\sf P}^{\psi}_{N,2,1}=1$. ${\EuScript P}_{N,2,1}$ may be realized as a union of three (trivial) fibre bundles 
$$
{\EuScript P}_{N,2,1}=\bigcup_{i=1}^3{\EuScript P}_{N,2,1}^{(i)}
\eqno(B.4)
$$
where each component ${\EuScript P}_{N,2,1}^{(i)}$ is homeomorphic to the Cartesian product ${\EuScript N}\times [0,1]$.   
\bigbreak
{\bf Appendix C: Construction of Parametric Diagonal Elements of Representable 1-Density Matrices with MATHEMATICA  Package }
\bigbreak
Matrix $H_{\psi}^{(1)}$ was constructed using the following code:  
\begin{multline*}
\\
\shoveleft {In[1]:=<< \rm LinearAlgebra\mbox{\textasciigrave}Orthogonalization\mbox{\textasciigrave}}\\
\shoveleft {In[2]:=\rm \{u1,u2,u3,u4,u5,u6\}=}\\
\qquad \rm GramSchmidt[\{\{0.8,0.4,0.3,0.2,0.1,0.4\},\{1,0,1,0,0,0\}\\
\{0, 1, 0, 1, 0, 0\}, \{0, 0, 1, 0, 1, 0\}, \{1, 0, 0, 0, 0, 1\},\qquad\\ 
\{-1, 1, 1, -1, 1, 1\}\}];\qquad\qquad\qquad\qquad\qquad\qquad\ \ \\
\shoveleft {In[3]:= \rm u1 = Outer[Times, u1, u1]; u2 = Outer[Times, u2, u2];}\\
\rm u3 = Outer[Times, u3, u3]; u4 = Outer[Times, u4, u4];\\
\rm  u5 = Outer[Times, u5, u5]; u6 = Outer[Times, u6, u6];\\
\qquad \rm H = -10*u1 + -5*u2 - 4*u3 - 3*u4 - 2*u5 - u6;\\
\\
\end{multline*}
Diagonal elements of 1-density matrix as functions of free parameters $x,y,z$ (see Eq.(\ref{vt})) were defined as
\begin{multline*}
\\
In[4]:=\rm \rho_{11}[x\_, y\_, z\_] := Block[\{m, v, h, ev, vec, ind, a, b, c, d, e, f, t\},\qquad \\
    \qquad\qquad\ \  \rm H = -10*u1 + -5*u2 - 4*u3 - 3*u4 - 2*u5 - u6;\\
    \rm { Xv=\begin{pmatrix}
     x+y&0&0&0&0&0\\
    0&x+z&0&0&0&0\\
    0&0&-y-z&0&0&0\\
    0&0&0&y+z&0&0\\
    0&0&0&0&-x-z&0\\
    0&0&0&0&0&-x-y
    \end{pmatrix};}\qquad \\
\\
\end{multline*}
\begin{multline*}
\\
    \rm h = H + Xv; \{ev, vec\} = Eigensystem[h];\quad\\
    \rm ind = Ordering[ev]; k = ind[[1]];\qquad\qquad\qquad\\
    \qquad\qquad\quad\rm a = vec[[k, 1]]; b = vec[[k, 2]]; c = vec[[k, 3]]; d = vec[[k, 4]];\\ 
    \qquad\qquad\quad\rm e = vec[[k, 5]]; f = vec[[k, 6]]; t = 0.5*(a^2 + b^2 + c^2)];\qquad\\
    \rm Plot[\rho_{11}[0, 0, z], \{z, -50, 50\}];\qquad\qquad\qquad\quad\\
    \\
\end{multline*}
Matrices  $H_{\psi}^{(2)}$ and $H_{\psi}^{(3)}$ were constructed in the analogous manner with the initial linearly independent vectors 
\begin{multline*}
(\frac{1}{2},\frac{1}{2},0,-\frac{1}{2},0,0),(0,1,1,0,0,0),(0,1,0,1,0,0),(0,0,1,0,-1,0),\\
(1,0,0,0,0,1),(-1,1,1,-1,1,1),\qquad\qquad\qquad\quad\qquad\qquad\qquad\quad\qquad\quad
\end{multline*} 
and 
\begin{multline*}
(\frac{4}{5},\frac{1}{5},0,0,0,0),(0,1,1,0,0,0),(0,1,0,1,0,0),(0,0,1,0,-1,0),\\
(1,0,0,0,0,1),(-1,1,1,-1,1,1),\qquad\qquad\qquad\quad\qquad\qquad\qquad\quad\qquad\quad
\end{multline*} 
respectively

\bigbreak

 \end{document}